# The stability of the 18-fold symmetry soft-matter quasicrystals

Zhi-Yi Tang[1] and Tian-You Fan[2]*

1 School of Computer Science and Technology, Beijing Institute of Technology, Beijing 100081, China

2 School of Physics, Beijing Institute of Technology, Beijing 100081, China

*Corresponding author, email: tyfan2013@163.com

**Abstract** Following our previous work this article reports a study on the stability of the 18-fold symmetry soft-matter quasicrystals, in which the extended free energy is a basis for the analysis that is similar to the study of the 12-fold symmetry quasicrystals. Due to the differences in structure between these two kinds of quasicrystals, their stabilities present some different characters to each other. Because there were no any results of the stability of 18-fold symmetry soft-matter quasicrystals reported yet, perhaps the present work is the first probe on the topic.

## 1. Introduction

Based on an extended free energy in Ref [1]the authors explored the stability of 12-fold symmetry soft-matter quasicrystals, which belongs to the first kind of soft-matter quasicrystals [2]. It is well-known the 18-fold symmetry soft-matter quasicrystals were observed in colloids [3].The 18-fold and possible discovered 7-, 9- and 14-fold symmetry quasicrystals belong to the second kind of soft-matter quasicrystals [4]. So the study on stability of 18-fold symmetry soft-matter quasicrystals is significant and different from that of the first kind of soft-matter quasicrystals. Based on the authors' detection, there were no any reported results of the stability for 18-fold symmetry soft-matter quasicrystals yet; perhaps the present work is the first probe on the topic.

Lack of research on stability of 18-fold symmetry soft-matter quasicrystals may be due to incomplete data of three-dimensional dynamics of the matter, for example, the constitutive law of the material is only for the two-dimensional case [5-7]. Although we derived the three-dimensional equations of dynamics of soft-matter quasicrystals, they just are for the first kind of ones [8] not including the second kind of ones, and the three-dimensional theory for 18-fold symmetry quasicrystals has just been obtained by the authors [9].

Lifshitz and Diamant[10] studied the stability of the 12-fold symmetry soft-matter quasicrystals in 2007 soon after the structure was observed. They are the pioneers of the work in which the basis is the effective free energy, that suggested by Lifshitz and Petrich [11]. Their subsequent studies and other researchers further developed the work [12-16].

Fan and his group have carried out the research on generalized dynamics of soft-matter quasicrystals [2,4,6-9] over the years, which provide some beneficial information to explore the physical background of the stability problem. A concept of extended free energy correlating the problem is found. This concept is very simple to help us to directly find the stability criterion, and the results are concise. Hence there is no need to carry any molecular dynamics simulation or other simulation. In Ref [1] the probe on the stability of 12-fold symmetry soft-matter quasicrystals was achieved, we obtained a very concise result. In this letter we report the study on stability of 18-fold symmetry soft-matter quasicrystals.

## 2. A brief review on some fundamental relations from the dynamics of second kind of soft-matter quasicrystals

We have mentioned in the previous section that the 18-fold and possible discovered 7-, 9- and 14-fold symmetry quasicrystals belong to the second kind of soft-matter quasicrystals [2], which are different from those of the first kind ones. They behave four elementary excitations such as phonons, first phasons, second phasons and fluid phonon. The relevant fields are phonon displacement field $\mathbf{u}(u_x,u_y,u_z)$ ,the first phason displacement field $\mathbf{v}(v_x,v_y)$ , the second phason displacement field $\mathbf{w}(w_x,w_y)$ and fluid phonon velocity field $\mathbf{V}(V_x,V_y,V_z)$ , respectively, in which the $z-$axis represents the 18- or 7-, 9-,14-axis.

We define the tensors of phonon train, phason strain and fluid phonon deformation rate as follows

$$\varepsilon_{ij}=\frac{1}{2}\left(\frac{\partial u_i}{\partial x_j}+\frac{\partial u_j}{\partial x_i}\right), v_{ij}=\frac{\partial v_i}{\partial x_j}, w_{ij}=\frac{\partial w_i}{\partial x_j}, \dot{\xi}_{ij}=\frac{1}{2}\left(\frac{\partial V_i}{\partial x_j}+\frac{\partial V_j}{\partial x_i}\right) \tag{1}$$

respectively, in which

$$x=x_1, y=x_2, z=x_3, \qquad i,j=1,2,3$$

and the corresponding constitutive law[2]

$$\left.\begin{aligned}
\sigma_{ij}&=C_{ijkl}\varepsilon_{kl}+r_{ijkl}v_{kl}+R_{ijkl}w_{kl}\\
\tau_{ij}&=T_{ijkl}v_{kl}+r_{klij}\varepsilon_{kl}+G_{ijkl}w_{kl}\\
H_{ij}&=K_{ijkl}w_{kl}+R_{klij}\varepsilon_{kl}+G_{klij}v_{kl}\\
p_{ij}&=-p\delta_{ij}+\sigma'_{ij}=-p\delta_{ij}+\eta_{ijkl}\dot{\xi}_{kl}
\end{aligned}\right\} \tag{2}$$

and inner energy densities

$$\left.\begin{aligned}
&U_{\mathrm{u}}=\frac{1}{2}C_{ijkl}\varepsilon_{ij}\varepsilon_{kl},\ U_{\mathrm{v}}=\frac{1}{2}T_{ijkl}v_{ij}v_{kl},\ U_{\mathrm{w}}=\frac{1}{2}K_{ijkl}w_{ij}w_{kl}\\
&U_{\mathrm{uv}}=r_{ijkl}\varepsilon_{ij}v_{kl}+r_{klij}v_{ij}\varepsilon_{kl}\\
&U_{\mathrm{uw}}=R_{ijkl}\varepsilon_{ij}w_{kl}+R_{klij}w_{ij}\varepsilon_{kl} \qquad i,j,k,l=1,2,3\\
&U_{\mathrm{vw}}=G_{ijkl}v_{ij}w_{kl}+G_{klij}w_{ij}v_{kl}
\end{aligned}\right\} \tag{3}$$

in which $C_{ijkl}$ denote the phonon elastic constants, $T_{ijkl}$ the first phason elastic constants, $K_{ijkl}$ the second phason elastic constants, $r_{ijkl}, r_{klij}$ and $R_{ijkl}, R_{klij}$ the phonon-first phason and phonon-second phason coupling elastic constants, $G_{ijkl}, G_{klij}$ the first-second phason coupling elastic constants [2, 4, 6-9] respectively. They are different from those for the 12-fold symmetry soft-matter quasicrystals given in Ref [1].

## 3. Extended free energy of the quasicrystal system in soft matter

We define the extended inner energy density

$$U_{ex}=\frac{1}{2}A\left(\frac{\delta\rho}{\rho_0}\right)^2+B\left(\frac{\delta\rho}{\rho_0}\right)\nabla\bullet\mathrm{u}+C\left(\frac{\delta\rho}{\rho_0}\right)\nabla\bullet\mathrm{v}+D\left(\frac{\delta\rho}{\rho_0}\right)\nabla\bullet\mathrm{w}+U_{el} \quad (4)$$

where the first term denotes energy density due to mass density variation, the quantity $\delta\rho/\rho_0$ describes the variation of the mass density, in which $\delta\rho=\rho-\rho_0$ and $\rho_0$ the initial mass density, according to our computation in the cases of transient response and flow past obstacle of soft-matter quasicrystals [7], $\delta\rho/\rho_0=10^{-4}\sim10^{-3}$ for soft-matter quasicrystals, which describes the fluid effect of the matter and is greater 10 order of magnitude than that of solid quasicrystals (in this sense we can consider for the solid quasicrystals the effect of $\delta\rho/\rho_0$ is very weak); the second term is one by mass density variation coupling phonons; the third and fourth represent that of mass density variation coupling first and second phasons, and $A,B,C,D$ the corresponding material constants, respectively. For quasicrystals $C$ and $D$ should be zero refer to [1]. Extending the work of Ref [1] the Hamiltonian

$$\left.\begin{aligned}
H&=H[\Psi(\mathrm{r},t)]\\
&=\int\frac{\mathrm{g}^2}{2\rho}d^d\mathrm{r}+\int\left[\frac{1}{2}A\left(\frac{\delta\rho}{\rho_0}\right)^2+B\left(\frac{\delta\rho}{\rho_0}\right)\nabla\bullet\mathrm{u}\right]d^d\mathrm{r}+F_{el}\\
&=H_{kin}+H_{\rho}+F_{el}\\
\mathrm{g}&=\rho\mathrm{V}
\end{aligned}\right\} \quad (5)$$

is discussed fully in Refs [1,2,4,12,13], in which $H_{kin}$ denotes the kinetic energy, $H_\rho$ the energy due to the variation of mass density, $F_{el}$ the elastic deformation energy consisting of contributed from phonons, phasons and phonon-phason coupling and phason-phason coupling, respectively, the detailed definition will be given by (6)-(8) in the following.

From the equations (2) one can obtain the total elastic free energy density

$$U_{el}=U_{\mathrm{u}}+U_{\mathrm{v}}+U_{\mathrm{w}}+U_{\mathrm{uv}}+U_{\mathrm{uw}}+U_{\mathrm{vw}} \quad (6)$$

which come from phonons, phasons, phonon-phason coupling and phason-phason coupling. And the elastic energy will be

$$F_{el}=F_{\mathrm{u}}+F_{\mathrm{v}}+F_{\mathrm{w}}+F_{\mathrm{uv}}+F_{\mathrm{uw}}+F_{\mathrm{vw}} \quad (7)$$

where $F_{\mathrm{u}},F_{\mathrm{v}},F_{\mathrm{w}},F_{\mathrm{uv}},F_{\mathrm{uw}},F_{\mathrm{vw}}$ represent the strain energies of phonons, first phasons, second phasons, phonon-first phason coupling, phonon-second phason coupling, and first phason-second phason coupling, respectively:

$$\left.\begin{aligned} F_{\mathrm{u}} &= \int \frac{1}{2} C_{ijkl} \varepsilon_{ij} \varepsilon_{kl} d^d \mathrm{r} \\ F_{\mathrm{v}} &= \int \frac{1}{2} T_{ijkl} v_{ij} v_{kl} d^d \mathrm{r} \\ F_{\mathrm{w}} &= \int \frac{1}{2} K_{ijkl} w_{ij} w_{kl} d^d \mathrm{r} \\ F_{\mathrm{uv}} &= \int \left( r_{ijkl} \varepsilon_{ij} v_{kl} + r_{klij} v_{ij} \varepsilon_{kl} \right) d^d \mathrm{r} \\ F_{\mathrm{uw}} &= \int \left( R_{ijkl} \varepsilon_{ij} w_{kl} + R_{klij} w_{ij} \varepsilon_{kl} \right) d^d \mathrm{r} \\ F_{\mathrm{vw}} &= \int \left( G_{ijkl} v_{ij} w_{kl} + G_{klij} w_{ij} v_{kl} \right) d^d \mathrm{r} \end{aligned}\right\} \tag{8}$$

The integrations are taken account on the domain.

Therefore, the problem of the second kind of quasicrystals is more complex than that of the first kind.

According to the thermodynamics, the extended free energy density is defined by

$$F_{ex} = U_{ex} - TS \tag{9}$$

where $U_{ex}$ the extended inner energy density defined by (4), and the $T$ absolute temperature and $S$ the entropy, respectively.

## Lemma

From (9) and (4)-(6), we have

$$S = -\frac{\partial F_{ex}}{\partial T}, \sigma_{ij} = \frac{\partial F_{ex}}{\partial \varepsilon_{ij}}, \tau_{ij} = \frac{\partial F_{ex}}{\partial v_{ij}}, H_{ij} = \frac{\partial F_{ex}}{\partial w_{ij}}, \delta^2 F_{ex} \geq 0 \tag{10}$$

in which the second to fourth equations are equivalent to the elastic constitutive law of the material for degrees of freedom of phonons, first and second phasons, and the last one is the stability condition of the matter, respectively. Because the formula (4) introduced an extended inner energy density, the variation principle containing in equations (10) is an extended or generalized variation, note that $C = D = 0$ in (4).

## 4. The positive definite nature of the rigidity matrix and the stability of the soft-matter quasicrystals with 18-fold symmetry

For point group $18\,mm$ of the 18-fold symmetry soft-matter quasicrystals the concrete constitutive law is as following

$$\left.\begin{aligned} \sigma_{xx} &= C_{11}\varepsilon_{xx} + C_{12}\varepsilon_{yy} + C_{13}\varepsilon_{zz} \\ \sigma_{yy} &= C_{12}\varepsilon_{xx} + C_{11}\varepsilon_{yy} + C_{13}\varepsilon_{zz} \\ \sigma_{zz} &= C_{13}\varepsilon_{xx} + C_{13}\varepsilon_{yy} + C_{33}\varepsilon_{zz} \\ \sigma_{yz} &= \sigma_{zy} = 2C_{44}\varepsilon_{yz} \\ \sigma_{zx} &= \sigma_{xz} = 2C_{44}\varepsilon_{zx} \\ \sigma_{xy} &= \sigma_{yx} = 2C_{66}\varepsilon_{xy} \end{aligned}\right\} \tag{11a}$$

$$\left.\begin{aligned}
\tau_{xx} &= T_1 v_{xx} + T_2 v_{yy} + G\left(w_{xx} - w_{yy}\right) \\
\tau_{yy} &= T_2 v_{xx} + T_1 v_{yy} + G\left(w_{xx} - w_{yy}\right) \\
\tau_{xy} &= T_1 v_{xy} - T_2 v_{yx} - G\left(w_{xy} + w_{yx}\right) \\
\tau_{yx} &= T_1 v_{yx} - T_2 v_{xy} + G\left(w_{xy} + w_{yx}\right) \\
\tau_{xz} &= T_3 v_{xz}, \qquad \tau_{yz} = T_3 v_{yz}
\end{aligned}\right\} \tag{11b}$$

$$\left.\begin{aligned}
H_{xx} &= K_1 w_{xx} + K_2 w_{yy} + G\left(v_{xx} + v_{yy}\right) \\
H_{yy} &= K_2 w_{xx} + K_1 w_{yy} - G\left(v_{xx} + v_{yy}\right) \\
H_{xy} &= K_1 w_{xy} - K_2 w_{yx} - G\left(v_{xy} - v_{yx}\right) \\
H_{yx} &= K_1 w_{yx} - K_2 w_{xy} - G\left(v_{xy} - v_{yx}\right) \\
H_{xz} &= K_3 w_{xz}, \qquad H_{yz} = K_3 w_{yz}
\end{aligned}\right\} \tag{11c}$$

$$\left.\begin{aligned}
p_{xx} &= -p + 2\eta\dot{\xi}_{xx} - \frac{2}{3}\eta\dot{\xi}_{kk} \\
p_{yy} &= -p + 2\eta\dot{\xi}_{yy} - \frac{2}{3}\eta\dot{\xi}_{kk} \\
p_{zz} &= -p + 2\eta\dot{\xi}_{zz} - \frac{2}{3}\eta\dot{\xi}_{kk} \\
p_{yz} &= p_{zy} = 2\eta\dot{\xi}_{yz} \\
p_{zx} &= p_{xz} = 2\eta\dot{\xi}_{zx} \\
p_{xy} &= p_{yx} = 2\eta\dot{\xi}_{xy} \\
\dot{\xi}_{kk} &= \dot{\xi}_{xx} + \dot{\xi}_{yy} + \dot{\xi}_{zz}
\end{aligned}\right\} \tag{11d}$$

the phonon-phason coupling constants $r_{ijkl} = r_{klij} = R_{ijkl} = R_{klij} = 0$ due to the decoupling between phonons and phasons, i.e.,

$$U_{\text{uv}} = r_{ijkl}\varepsilon_{ij}v_{kl} + r_{klij}v_{ij}\varepsilon_{kl} = 0$$

$$U_{\text{uw}} = R_{ijkl}\varepsilon_{ij}w_{kl} + R_{klij}w_{ij}\varepsilon_{kl} = 0$$

for this type of quasicrystals.

As well as the conventional inner energy density, between stress tensor and strain tensor there is an elastic rigidity matrix, for the extended inner energy density (4) there is an extended rigidity matrix such as

$M_1 =$

$$
\begin{pmatrix}
A & B & B & B & & & & & & & & & & & & & & & \\
B & C_{11} & C_{12} & C_{13} & & & & & & & & & & & & & & & \\
B & C_{12} & C_{11} & C_{13} & & & & & & & & & & & & & & & \\
B & C_{13} & C_{13} & C_{33} & & & & & & & & & & & & & & & \\
 & & & & 2C_{44} & & & & & & & & & & & & & & \\
 & & & & & 2C_{44} & & & & & & & & & & & & & \\
 & & & & & & 2C_{66} & & & & & & & & & & & & \\
 & & & & & & & T_1 & T_2 & & & & & G & -G & & & & \\
 & & & & & & & T_2 & T_1 & & & & & G & -G & & & & \\
 & & & & & & & & & T_3 & & & & & & & & & \\
 & & & & & & & & & & T_1 & & -T_2 & & & & -G & & -G \\
 & & & & & & & & & & & T_3 & & & & & & & \\
 & & & & & & & & & & -T_2 & & T_1 & & & & G & & G \\
 & & & & & & & G & G & & & & & K_1 & K_2 & & & & \\
 & & & & & & & -G & -G & & & & & K_2 & K_2 & & & & \\
 & & & & & & & & & & & & & & & K_3 & & & \\
 & & & & & & & & & & -G & & G & & & & K_1 & & -K_2 \\
 & & & & & & & & & & & & & & & & & K_3 & \\
 & & & & & & & & & & -G & & G & & & & -K_2 & & K_1
\end{pmatrix} \tag{12}
$$

Due to the condition in (10)

$$\delta^2 F_{ex} \geq 0 \tag{13}$$

this non-negative condition of the second order variation of the extended free energy density functional requires the extended rigidity matrix must be positive definite. We have the theorem for describing the stability of the soft-matter quasicrystals with 18-fold symmetry as follows:

## Theorem

Under the condition (4), the validity of variation (13) is equivalent to the positive definite nature of matrix (12) and leads to

$$
\left.
\begin{aligned}
&A > 0,\ C_{11} - C_{12} > 0,\ C_{44} > 0,\ A(C_{11}C_{33} + C_{12}C_{33} - 2C_{13}^2) - B^2\left(C_{11} + C_{12} - 4C_{13} + 2C_{33}\right) > 0, \\
&K_1 + K_2 > 0,\ T_1 - T_2 > 0,\ \left(K_1 - K_2\right)\left(T_1 + T_2\right) - 4G^2 > 0,\ K_1 - K_2 > 0,\ T_3 > 0,\ K_3 > 0
\end{aligned}
\right\} \tag{14}
$$

The proof of the theorem is in straight forward manner, so it is omitted.

When the conditions (14) are satisfied, the 18-fold symmetry soft-matter quasicrystals are stable. This stability takes into account of the effects of fluid, fluid coupling phonons, phonons, phasons and the coupling between the first and second phasons of 18-fold symmetry soft-matter quasicrystals, more exactly speaking the stability depends upon these material constants. These constants can be measured by experiments which are similar to that in crystallography [17] and solid quasicrystallography [1]. These present simplicity and intuitive character of the complexity of stability of soft-matter quasicrystals. This shows the substantive nature of the stability of soft-matter quasicrystals. Naturally it explores the structure of the matter, because it comes from the constitutive law (11), which is the result of the symmetry of the structure---i.e., the result obtained by theory of group and group representation of the quasicrystals, refer to [2,4, 6-8].

## 4. Comparison and examination

It is well-known that the 18-fold symmetry quasicrystals in soft matter belong to a type of two-dimensional quasicrystals, and in which the phonon field structure presents the character of the hexagonal crystals [2,4,6-8].

For the special case, as phason field is absent, i.e.,

$$K_1 = K_2 = T_1 = T_2 = 0, G = 0 \tag{15}$$

and at the same time one takes $B = 0$ and $A$ being any positive value, then (14) reduces to Cowley[17]

$$C_{11} - C_{12} > 0,\ C_{44} > 0,\ C_{11}C_{33} + C_{12}C_{33} - 2C_{13}^2 > 0 \tag{16}$$

Here constant $A$ can be taken arbitrary small, due to $\left(\frac{\delta\rho}{\rho_0}\right)^2 \sim 10^{-6}$ (because of $\frac{\delta\rho}{\rho_0} \sim 10^{-3}$, according to our computation for soft-matter quasicrystals), so $A\left(\frac{\delta\rho}{\rho_0}\right)^2$ is very small, this may be understood that if the fluid effect is quite weak and no phason field, the stability condition (14) is reduced to that of hexagonal crystal system, the latter was derived by Cowley[17].

While for another case, i.e., in the solid quasicrystals of 18-fold symmetry, although this type of solid quasicrystals has not been observed by experiments so far,the possible structure can be estimated from group theory, then (14) reduces to

$$\left.\begin{array}{l} C_{11} - C_{12} > 0,\ C_{44} > 0,\ C_{11}C_{33} + C_{12}C_{33} - 2C_{13}^2 > 0, \\ K_1 + K_2 > 0,\ T_1 - T_2 > 0,\ (K_1 - K_2)(T_1 + T_2) - 4G^2 > 0,\ K_1 - K_2 > 0,\ T_3 > 0,\ K_3 > 0 \end{array}\right\} \tag{17}$$

under condition $B$=0 and $A$ being any positive value, the inequalities (17) are the stability condition of solid quasicrystals of 18-fold symmetry, which are derived by the authors and explore the correctness of the dynamics and thermodynamics of soft-matter quasicrystals once again.

The stability is connected with the positive definite nature of the mathematical structure of the dynamics, this is also useful for the numerical solution (e.g. the finite element method), which will be discussed in our another paper.

The stability is correlated to the phase transition, this is a more important problem. For crystals, Cowley [17] gave an analysis, and for quasicrystals, it requires to carry out to continue the probe.

## 5. Conclusion and discussion

The theorem can also be given in accordance with the first law of thermodynamics, by omitting details we have

$$\sigma_{ij} = \frac{\partial U_{ex}}{\partial \varepsilon_{ij}}, \tau_{ij} = \frac{\partial U_{ex}}{\partial v_{ij}}, H_{ij} = \frac{\partial U_{ex}}{\partial w_{ij}}, \delta^2 U_{ex} \geq 0 \tag{18}$$

where $U_{ex}$ is defined by (4), which is a quadratic form, the condition $\delta^2 U_{ex} \geq 0$ requires the matrix (12) must be positive definite , so leads to the results (14), i.e., the theorem holds.

In addition, we here should point out that in Ref [1], there was a type error, i.e.,

in (16) of Ref [1]

$$T = \frac{\partial U_{ex}}{\partial S}$$

should be removed.

By a complete different approach compared with Refs [10-16] this report discussed the stability of soft-matter quasicrystals, which is directly based on the thermodynamics with the help of generalized dynamics of the matter, and offered some quantitative and very concise results, the stability depends only upon the material constants, which can be measured by experiments. The correctness and precision of the theoretical prediction is examined by results of crystals and solid quasicrystals in qualitatively as well as quantitatively (refer to (16) and (17)). In the examination through the crystals and solid quasicrystals, we find that the constant $A$ can be arbitrary positive number but does not equal to zero, this shows the soft matter state cannot be reduced to any solid state phase, so the constant presents an important meaning for soft matter. The introducing of the constant $A$ in studying the generalized dynamics of soft-matter quasicrystals can refer to Refs [1,2,4,12,13] in detail. The thermodynamics of the matter shows the constant $A$ becomes more important.

For possible soft-matter quasicrystals with 7-, 9- and 14-fold symmetry the stabilities are similar to that given by (14), we will report in other case.

The work on stability of 12- and 18-fold symmetry quasicrystals found that

1) The stability of 12-fold symmetry quasicrystals depends upon the fluid effect, describing by constant $A$, fluid coupling to phonons describing by constant $B$, phonons describing by $C_{ijkl}$ and phasons describing by $K_{ijkl}$ ;

2) The stability of 18-fold symmetry quasicrystals depends upon the fluid effect, fluid coupling to phonons, phonons and phasons (which are similar to those in[1]) and coupling between first and second phasons describing by $G_{ijkl}$ ;

3) These constants can be measured by experiments;

4) The differences between those of 12- and 18-fold symmetry quasicrystals lie in their different structures, this is explored by the constitutive laws (i.e., the equations (9) in Ref [1] and (11) in this paper), which are determined by the theory of group representation.

5) Due to the combination between thermodynamics and generalized dynamics of the matter, approach determining the stability developed by Ref [1] and this paper presents systematic, direct and simple features, it is without and needed not using any simulation treatment, and the results obtained present bright physical meaning, whose correctness, of course, has to be verified by experiments further.

**Acknowledgement** The work is supported by the National Natural Science Foundation of China through the grant 11272053. Tian-You Fan thanks Prof R Lifshitz of Tel Aviv University in Israel for presenting the electronic copy of Refs [10,12]. Zhi-Yi Tang is also grateful to the support in part of the National Natural Science Foundation of China through the grant 11871098.